\documentstyle[12pt]{article}

\catcode `\@=11 \@addtoreset{equation}{section}
\def\theequation{\arabic{section}.\arabic{equation}}
\catcode `\@=12

\newcommand{\be}{\begin{equation}}
\newcommand{\en}{\end{equation}}
\newcommand{\bea}{\begin{eqnarray}}
\newcommand{\ena}{\end{eqnarray}}
\newcommand{\beano}{\begin{eqnarray*}}
\newcommand{\enano}{\end{eqnarray*}}
\newcommand{\bee}{\begin{enumerate}}
\newcommand{\ene}{\end{enumerate}}

\newcommand{\Z}{Z \!\!\!\!\! Z}

\newcommand{\Id}{1\!\!1}

\newcommand{\E}{{\cal E}}

\begin{document}

\thispagestyle{empty}

\vspace*{1cm}

\begin{center}
{\Large \bf The stochastic limit in the analysis of the open BCS model}\footnote{This
paper is dedicated with all my love to my sweet father}   \vspace{2cm}\\

{\large F. Bagarello}
\vspace{3mm}\\
  Dipartimento di Matematica ed Applicazioni,
Facolt\`a di Ingegneria, Universit\`a di Palermo, \\Viale delle Scienze, I-90128  Palermo, Italy\\
e-mail: bagarell@unipa.it
\vspace{4mm}\\

\end{center}

\vspace*{2cm}

\begin{abstract}
\noindent In this paper we show how the perturbative procedure
known as {\em stochastic limit} may be useful in the analysis of
the Open BCS model  discussed by Buffet and Martin as a spin
system interacting with a fermionic reservoir. In particular we
show how the same values of the critical temperature and of the
order parameters can be found with a significantly simpler
approach.
\end{abstract}

\vspace{2cm}

 \vfill

\newpage

\section{Introduction}

In this paper we analyze the Open BCS model as given in
\cite{bm,martin} using the techniques of the {\em stochastic limit
approach} (SLA), which is described in many details in the
monograph \cite{book}. Instead of considering a fermionic
reservoir, as the authors do in \cite{bm,martin} (following the
original suggestion contained in \cite{hl} which allow to avoid
dealing with unbounded operators), we will consider here a bosonic
thermal bath. This choice is made to try to stay closer to the
real physical world, where the reservoir is bosonic. This means
that some of our formulas are only formal, but they can be made
rigorous with just a little effort, using, for instance, the same
framework for unbounded operators developed in \cite{bagjmp} and
references therein. We will comment again on this aspect of our
model in the next section.

The main outcome of this paper is that the same values of the
critical temperature and of the order parameters can be found
using the SLA, in a significantly simpler way, as we will show in
Section III. This simplification allows us to focus our attention
on some aspects of the model which could appear not so clearly
using the standard technique. This is what has been already
observed in other physical applications: for instance, in
\cite{baglaser}, we used the SLA to explore in details some
relations between different models of matter interacting with the
radiation, as the Hepp-Lieb and the Alli-Sewell models. Also, in
\cite{bagacc}, the SLA was used in connection with the fractional
quantum Hall effect, giving some interesting results. Other
applications are contained in \cite{book} and references therein.

The paper is organized as follows:

in the next section we introduce the model and compute its
generator  using the SLA together with a semiclassical
approximation, already introduced  in \cite{martin}, useful to
obtain the free evolution of the matter operators;

in Section III we write the equations of motion for some
macroscopic variables of the matter and we recover the same
results as in \cite{bm};

\noindent our conclusions are contained in Section IV, while the
Appendix is devoted to review some facts concerning the SLA,
useful to keep the paper self contained.

\section{The Physicals Model and its stochastic limit}

Our model consists of two main ingredients, the {\em system},
which is described by spin variables, and the {\em reservoir},
which is given in terms of bosonic operators. It is contained in a
box of volume $V=L^3$, with $N$ lattice sites. We define,
following \cite{bm,martin} \be H_N^{(sys)}=\tilde\epsilon
\sum_{j=1}^N\sigma_j^0-\frac{g}{N}\sum_{i,j=1}^N\sigma_i^+\sigma_j^-,
\label{21} \en where the indexes $i,j$ represent the discrete
values of the momentum that an electron in a fixed volume can
have, $\sigma_j^+$ creates a Cooper pair with given momentum while
$\sigma_j^-$ annihilates the same pair, $\tilde\epsilon$ is the
energy of a single electron and $-g<0$ is the interaction close to
the Fermi surface. As we can see, only the $\pm$ component of the
spin, that is the $x,y$ components, have a mean field interaction,
while the $z$ component interacts with a constant external
magnetic field. The algebra of the Pauli matrices is given by \be
[\sigma_i^+,\sigma_j^-]=\delta_{ij}\sigma_i^0,\hspace{1cm}
[\sigma_i^\pm,\sigma_j^0]=\mp 2\delta_{ij}\sigma_i^\pm.
\label{22}\en We will use the following realization of these
matrices:
$$
\sigma^0\equiv\sigma^z=\left(
\begin{array}{cc}
1 & 0   \\
0 & -1  \\
\end{array}
\right), \hspace{5mm} \sigma^+=\left(
\begin{array}{cc}
0 & 1   \\
0 & 0  \\
\end{array}
\right), \hspace{5mm} \sigma^-=\left(
\begin{array}{cc}
0 & 0   \\
1 & 0  \\
\end{array}
\right).
$$
If we now define the following operators, \ \be
S_N^\alpha=\frac{1}{N}\sum_{i=1}^N\sigma_i^\alpha, \hspace{5mm}
R_N=S_N^+S_N^-=R_N^\dagger, \label{23}\en $H_N^{(sys)}$ can be
simply written as $H_N^{(sys)}=N(\tilde\epsilon S_N^0-gR_N)$ and
it is easy to check that the following commutation rules hold:
$$
[S_N^0,R_N]=[H_N^{(sys)},R_N]=[H_N^{(sys)},S_N^0]=0,
$$
for any given $N>0$. It is also worth noticing that the intensive
operators $S_N^\alpha$ are all bounded by 1 in the operator norm,
and that the commutators $[S_N^\alpha,\sigma_j^\beta]$ go to zero
in norm as $\frac{1}{N}$ when $N\rightarrow\infty$, for all $j,
\alpha$ and $\beta$.

\vspace{3mm}

As we have already mentioned in the Introduction, we consider here
a realistic bosonic reservoir, so that some of the following
formulas must be understood to be {\em formal}. However, using for
instance the same algebraic framework discussed in \cite{bagjmp}
for some different spin-bosons models, or replacing the bosonic
operators with their smeared versions, everything can be made
rigorous. We avoid here this useless complication, since it would
make all the treatment much more complicated, hiding in this way
our main results.

Our construction of the reservoir follows the same steps given in
\cite{martin}, but for the commutation rules. We introduce here as
many bosonic modes $a_{\vec p,j}$ as lattice sites are present in
$V$. This means that $j=1,2,...,N$. $\vec p$ is the value of the
momentum of the j-th boson which, if we impose periodic boundary
condition on the wave functions, has necessarily the form $\vec
p=\frac{2\pi}{L}\vec n$, where $\vec n=(n_1,n_2,n_3)$ with
$n_j\in\Z$. These operators satisfy the following CCR, \be
[a_{\vec p,i},a_{\vec q,j}]=[a_{\vec p,i}^\dagger,a_{\vec
q,j}^\dagger]=0, \hspace{5mm} [a_{\vec p,i},a_{\vec
q,j}^\dagger]=\delta_{ij}\delta_{\vec p\,\vec q} \label{24}\en and
their free dynamics is given by \be
H_N^{(res)}=\sum_{j=1}^N\sum_{\vec p\in\Lambda_N}\epsilon_{\vec
p}\, a_{\vec p,j}^\dagger a_{\vec p,j}. \label{25} \en Here
$\Lambda_N$ is the set of values which $\vec p$ may take,
according to the previous remark: $\Lambda_N=\{\vec
p=\frac{2\pi}{L}\vec n,\, \vec n\in\Z^3\}$. It is useful to stress
that the energy of the different bosons is clearly independent of
the lattice site: $\epsilon_{\vec p}=\frac{\vec p^2}{2m} =
\frac{4\pi^2(n_1^2+n_2^2+n_3^2)}{2mL^2}$.

The form of the interaction between reservoir and system is
assumed to be of the following form: \be
H_N^{(I)}=\sum_{j=1}^N(\sigma_j^+a_j(f)+h.c.),\label{26}\en where
we have introduced $a_j(f)=\sum_{\vec p\in\Lambda_N}a_{\vec
p,j}f(\vec p)$, $f$ being a given test function which will be
asked to satisfy some extra conditions, see equation (\ref{221})
below and the related discussion. We would like to stress that, in
order to keep the notation reasonably simple, we will not use the
tensor product symbol along this paper whenever the meaning of the
symbols is clear.

The finite volume open system is now described by the following
hamiltonian, \be H_N=H_N^0+\lambda H_N^{(I)}, \mbox{ where }
H_N^0=H_N^{(sys)}+H_N^{(res)} \label{27}\en and $\lambda$ is the
coupling constant.

The first step in the SLA is the computation of the free evolution
of the interaction hamiltonian: \be
H_N^{(I)}(t)=e^{iH_N^0t}H_N^{(I)}e^{-iH_N^0t}=\sum_{j=1}^N(e^{iH_N^{(sys)}t}\sigma_j^+
e^{-iH_N^{(sys)}t}e^{iH_N^{(res)}t}a_j(f)
e^{-iH_N^{(res)}t}+h.c.). \label{28}\en The computation of the
part of the reservoir is trivial and produces
$$e^{iH_N^{(res)}t}a_j(f) e^{-iH_N^{(res)}t}=a_j(fe^{-it\epsilon
}),$$ where $a_j(fe^{-it\epsilon })=\sum_{\vec
p\in\Lambda_N}a_{\vec p,j}f(\vec p)e^{-it\epsilon_{\vec p}}$. This
is an easy consequence of the CCR (\ref{24}). The free evolution
of the spin operators is more difficult and its expression can be
found in \cite{bm,martin}, for instance. Here it is shown how to
obtain the time evolution in a {\em semiclassical} approximation,
i.e., when the free time evolution of the intensive operators
$S_N^\alpha$ is replaced by its limit (taken in the strong
topology restricted to a certain family of relevant vectors,
\cite{bagmor}).

The differential equations of motion for the spin variables are
\be \left\{
\begin{array}{ll}
\frac{d\sigma_j^+(t)}{dt} = 2i\tilde\epsilon \sigma_j^+(t)+igS_N^+(t)\sigma_j^0(t)  \\
\frac{d\sigma_j^0(t)}{dt} = 2ig(\sigma_j^+(t)S_N^-(t)-\sigma_j^-(t)S_N^+(t)). \\
\end{array}
\right. \label{29} \en where we have called, with a little abuse
of language which is quite useful to maintain the notation simple,
$\sigma_j^\alpha(t)= e^{iH_N^{(sys)}t}\sigma_j^\alpha
e^{-iH_N^{(sys)}t}$. In fact, to be more precise, instead of
$\sigma_j^\alpha(t)$, we should write
$\sigma_{j,N}^{\alpha,free}(t)$, to stress the fact that
$e^{iH_N^{(sys)}t}\sigma_j^\alpha e^{-iH_N^{(sys)}t}$ only
produces the free evolution of $\sigma_j^\alpha$, i.e., the
evolution without any reservoir, and for $N$ fixed. Moreover, in
(\ref{29}) we have introduced
$S_N^\alpha(t)=e^{iH_N^{(sys)}t}S_N^\alpha e^{-iH_N^{(sys)}t}=
\frac{1}{N}\sum_{j=1}^N e^{iH_N^{(sys)}t}\sigma_j^\alpha
e^{-iH_N^{(sys)}t}= \frac{1}{N}\sum_{j=1}^N \sigma_j^\alpha(t)$.

Let us now call $S^\alpha={\mathcal F}-strong
\lim_{N\rightarrow\infty}S^\alpha_N$. The proof of the existence
of this limit (together with all its powers) may be found in
\cite{bagmor} and references therein. We can now take the sum over
$j=1,2,...,N$ of (both sides of) the equations in (\ref{29}),
divide the result by $N$, and consider the ${\mathcal F}-strong
\lim_{N\rightarrow\infty}$ of the equations obtained in this way.
We find that $\dot S^0(t)=0$ and $\dot
S^+(t)=i(2\tilde\epsilon+gS^0(t))S^+(t)$. These equations can be
easily solved: $S^0(t)=S^0=(S^0)^\dagger$ and
$S^+(t)=S^+e^{i(2\tilde\epsilon+gS^0)t}$. Of course $S^-(t)=
(S^+(t))^\dagger$. The system (\ref{29}) gives now, if we replace
$S_N^\alpha(t)$ with its ${\mathcal F}-strong$ limit
$S^\alpha(t)$, \be \left\{
\begin{array}{ll}
\frac{d\sigma_j^+(t)}{dt} = 2i\tilde\epsilon \sigma_j^+(t)+igS^+(t)\sigma_j^0(t)  \\
\frac{d\sigma_j^0(t)}{dt} = 2ig(\sigma_j^+(t)S^-(t)-\sigma_j^-(t)S^+(t)). \\
\end{array}
\right. \label{29bis} \en This system is called the {\em
semiclassical} approximation of (\ref{29}), and it can be
explicitly solved using, for instance, the Laplace transform
technique. The computation is rather long and we omit here all the
details, which can be found in \cite{bm,martin} . Also, since only
$\sigma_j^+(t)$ appear in (\ref{28}), together with its hermitian
conjugate, we give here only the result we need. We have \be
\sigma_j^+(t)=e^{i\nu t}\rho_0^j+e^{i(\nu+\omega)
t}\rho_+^j+e^{i(\nu-\omega) t}\rho_-^j, \label{210}\en where we
have defined the following operators \be \left\{
\begin{array}{ll}
\rho_0^j = \frac{g^2 S^+}{\omega^2}\left(2S^-\sigma_j^++S^0\sigma_j^0+2S^+\sigma_j^-\right)  \\
\rho_+^j = \frac{g
S^+}{\omega^2}\left(gS^-\frac{\omega-gS^0}{\omega+gS^0}\sigma_j^++
\frac{\omega-gS^0}{2}\sigma_j^0-gS^+\sigma_j^-\right) \\
\rho_-^j = \frac{g S^+}{\omega^2}
\left(gS^-\frac{\omega+gS^0}{\omega-gS^0}\sigma_j^+-
\frac{\omega+gS^0}{2}\sigma_j^0-gS^+\sigma_j^-\right), \\
\end{array}
\right. \label{211} \en and the following quantities \be
\omega=g\sqrt{(S^0)^2+4S^+S^-},\: \nu=2\tilde\epsilon+gS^0.
\label{212}\en Defining further \be \nu_\alpha(\vec
p)=\nu-\epsilon_{\vec p}+\alpha\omega, \label{212bis}\en
 where $\alpha$ takes the values $0$, $+$ and
$-$, the operator $H_N^{(I)}(t)$ in (\ref{28}) becomes \be
H_N^{(I)}(t)=\sum_{j=1}^N\sum_{\alpha=0,\pm}\left(\rho_\alpha^ja_j(fe^{it\nu_\alpha})+h.c\right).
\label{213}\en

{\bf Remark:--} It may be worth remarking that we would have
obtained exactly this free time evolution even for a fermionic
reservoir, since CCR and CAR produce the same free time evolution
for both the annihilation and the creation operators. From this
point of view, the difference between a fermionic and a bosonic
thermal bath appears really only  a minor aspect of the model.

The next step in the SLA consists in computing the following
quantity \be
I_\lambda(t)=\left(-\frac{i}{\lambda}\right)^2\int_0^t dt_1
\int_0^{t_1}dt_2\,\omega_{tot}\left(H_N^{(I)}(\frac{t_1}{\lambda^2})H_N^{(I)}(\frac{t_2}{\lambda^2})\right),\label{214}\en
and its limit for $\lambda$ going to zero. Here the state
$\omega_{tot}$ is the following product state
$\omega_{tot}=\omega_{sys}\,\omega_\beta$, where $\omega_{sys}$ is
a state of the system, while $\omega_\beta$ is a state of the
reservoir, which we will take to be a KMS state corresponding to
an inverse temperature $\beta=\frac{1}{kT}$. It is convenient here
to use the so-called {\em canonical representation of thermal
states}, \cite{book}, which is sketched in the Appendix. Then we
introduce two sets of mutually commuting bosonic operators
$\{c_{\vec p,j}^{(\gamma)}\}$, $\gamma=a,b$, as follows: \be
a_{\vec p,j}=\sqrt{m(\vec p)}\,c_{\vec p,j}^{(a)}+\sqrt{n(\vec
p)}\,c_{\vec p,j}^{(b),\dagger},\label{215}\en where \be m(\vec
p)=\omega_\beta(a_{\vec p,j}a_{\vec
p,j}^\dagger)=\frac{1}{1-e^{-\beta\epsilon_{\vec p}}},
\hspace{1cm}n(\vec p)=\omega_\beta(a_{\vec p,j}^\dagger a_{\vec
p,j})=\frac{e^{-\beta\epsilon_{\vec p}}}{1-e^{-\beta\epsilon_{\vec
p}}}.\label{215bis}\en The operators $c_{\vec p,j}^{(\alpha)}$
satisfy the following commutation rules \be [c_{\vec
p,j}^{(\alpha)},{c_{\vec
q,k}^{(\gamma)}}^\dagger]=\delta_{jk}\delta_{\vec p\,\vec
q}\delta_{\alpha\gamma}, \label{216}\en while all the other
commutators are trivial. Furthermore, we introduce  the vacuum of
the operators $c_{\vec p,j}^{(\alpha)}$, $\Phi_0$: \be c_{\vec
p,j}^{(\alpha)}\Phi_0=0, \hspace{1cm}\forall \vec p\in\Lambda_N,\:
j=1,..N,\: \alpha=a,b. \label{219}\en Finally, if we define
$f_m(\vec p)=\sqrt{m(\vec p)}f(\vec p)$ and $f_n(\vec
p)=\sqrt{n(\vec p)}f(\vec p)$, we get \be
a_j(fe^{it\nu_\alpha})=c_j^{(a)}(f_me^{it\nu_\alpha}) +
{c_j^{(b)}}^\dagger(f_ne^{it\nu_\alpha}),\label{217}\en where we
have used the usual following notation $c_j^{(a)}(g)=\sum_{\vec
p\in\Lambda_N}c_{\vec p,j}^{(a)}g(\vec p)$ and
${c_j^{(b)}}^\dagger(g)= \sum_{\vec p\in\Lambda_N}{c_{\vec
p,j}^{(b)}}^\dagger g(\vec p)$\footnote{It may be worth noticing
that both $c_j^{(\gamma)}(f)$ and ${c_j^{(\gamma)}(f)}^\dagger$
are linear in their argument $f$}. Therefore we have \be
H_N^{(I)}(t)=\sum_{j=1}^N\sum_{\alpha=0,\pm}\left\{\rho_\alpha^j\left(c_j^{(a)}(f_me^{it\nu_\alpha})
+ {c_j^{(b)}}^\dagger(f_ne^{it\nu_\alpha})\right)+h.c\right\},
\label{218}\en and the KMS state $\omega_\beta$ can be represented
as the following vector state, as in a GNS-like representation:
\be \omega_\beta(X_r)=<\Phi_0, X_r \Phi_0>, \label{220}\en for any
observable of the reservoir, $X_r$, since $\omega_\beta$ is a
gaussian state, \cite{book}. This fact, together with (\ref{219})
and with the commutation rules (\ref{216}), simplifies the
computation of the two point function
$\omega_{tot}\left(H_N^{(I)}(\frac{t_1}{\lambda^2})H_N^{(I)}(\frac{t_2}{\lambda^2})\right)$,
which, after some algebraic computations, produces
$$
\omega_{tot}\left(H_N^{(I)}(\frac{t_1}{\lambda^2})H_N^{(I)}(\frac{t_2}{\lambda^2})\right)=\sum_{j=1}^N
\sum_{\alpha,\beta=0,\pm}\sum_{\vec
p\in\Lambda_N}\{\omega_{sys}(\rho_\alpha^j{\rho_\beta^j}^\dagger)|f_m(\vec
p)|^2 e^{i\frac{t_1}{\lambda^2}\nu_\alpha(\vec p)}
e^{-i\frac{t_2}{\lambda^2}\nu_\beta(\vec p)}+$$ \vspace{-4mm}
$$+\, \omega_{sys}({\rho_\alpha^j}^\dagger \rho_\beta^j)|f_n(\vec
p)|^2 e^{-i\frac{t_1}{\lambda^2}\nu_\alpha(\vec p)}
e^{+i\frac{t_2}{\lambda^2}\nu_\beta(\vec p)} \}.
$$
Since we are interested to the limit $\lambda\rightarrow 0$ of
$I_\lambda(t)$ we need to impose some conditions on the test
function $f(\vec p)$, \cite{book}. In particular, we will require
that the following integral exists finite: \be
\int_{-\infty}^0\,d\tau \sum_{\vec p\in\Lambda_N}|f_r(\vec
p)|^2e^{\pm i\tau\nu_\alpha(\vec p)}<\infty,\label{221}\en where
$f_r(\vec p)$ is $f_m(\vec p)$ or $f_n(\vec p)$ and
$\nu_\alpha(\vec p)$ is given in (\ref{212bis}). Under this
assumption we find that \be I(t)=\lim_{\lambda\rightarrow
0}I_\lambda(t)=-t\sum_{j=1}^N\sum_{\alpha=0,\pm}\left\{\omega_{sys}
(\rho_\alpha^j{\rho_\alpha^j}^\dagger)\Gamma_\alpha^{(a)}+
\omega_{sys}({\rho_\alpha^j}^\dagger
\rho_\alpha^j)\Gamma_\alpha^{(b)}\right\}, \label{222}\en where
the two complex quantities \be
\Gamma_\alpha^{(a)}=\int_{-\infty}^0\,d\tau \sum_{\vec
p\in\Lambda_N}|f_m(\vec p)|^2e^{- i\tau\nu_\alpha(\vec p)},
\hspace{5mm} \Gamma_\alpha^{(b)}=\int_{-\infty}^0\,d\tau
\sum_{\vec p\in\Lambda_N}|f_n(\vec p)|^2e^{i\tau\nu_\alpha(\vec
p)} \label{223}\en both exist because of the assumption
(\ref{221}).

To this  same result we could also arrive starting with the
following {\em stochastic limit hamiltonian} \be
H_N^{(sl)}(t)=\sum_{j=1}^N\sum_{\alpha=0,\pm}\left\{\rho_\alpha^j\left(c_{\alpha
j}^{(a)}(t) + {c_{\alpha j}^{(b)}}^\dagger(t)\right)+h.c\right\},
\label{224}\en where the operators $c_{\alpha j}^{(\gamma)}(t)$
are assumed to satisfy the following commutation rule, \be
[c_{\alpha j}^{(\gamma)}(t),{c_{\beta
k}^{(\mu)}}^\dagger(t')]=\delta_{jk}\,\delta_{\alpha
\beta}\,\delta_{\gamma \mu} \delta(t-t')\Gamma_\alpha^{(\gamma)},
\hspace{1.4cm}\mbox{for } t>t'. \label{225}\en

We mean that, as it is easily checked, the following quantity
$$J(t)=(-i)^2\int_0^t dt_1
\int_0^{t_1}dt_2\Omega_{tot}(H_N^{(sl)}(t_1)H_N^{(sl)}(t_2))$$
coincides with $I(t)$. Here $\Omega_{tot}= \omega_{sys}\,
\Omega=\omega_{sys}\, <\Psi_0,\,\Psi_0>$, where $\Psi_0$ is the
vacuum of the operators $c_{\alpha j}^{(\gamma)}(t)$: $c_{\alpha
j}^{(\gamma)}(t)\Psi_0=0$ for all $\alpha, j, \gamma$ and $t$,
\cite{book}.

\vspace{2mm}

Following the SLA, \cite{book}, we now use $H_N^{(sl)}(t)$ to
compute the generator of the theory. In fact, this is the main
reason why this operator is introduced in the game. Let $X$ be an
observable of the system and $\Id_r$ the identity of the
reservoir. Its time evolution (after the stochastic limit is
taken) is $j_t(X\otimes \Id_r)=U_t^\dagger (X\otimes \Id_r) U_t$,
where $U_t$ is the wave operator satisfying the following
differential equation $\partial_tU_t=-iH_N^{(sl)}(t)U_t$, whose
adjoint is $\partial_tU_t^\dagger=iU_t^\dagger H_N^{(sl)}(t)$.

Then we find
$$
\partial_tj_t(X\otimes\Id_r)=iU_t^\dagger[H_N^{(sl)}(t),X\otimes\Id_r]U_t=$$
\vspace{-6mm} $$=iU_t^\dagger\sum_{j=1}^N\sum_{\alpha=0,\pm}
\left\{[\rho_\alpha^j,X](c_{\alpha j}^{(a)}(t)+{c_{\alpha
j}^{(b)}}^\dagger(t))+ [{\rho_\alpha^j}^\dagger,X]({c_{\alpha
j}^{(a)}}^\dagger(t)+c_{\alpha j}^{(b)}(t)) \right\}U_t
$$
Next we have to  normal order the formula above, i.e. to move to
the right all the annihilation operators $c_{\alpha
j}^{(\gamma)}(t)$ and to the left the creation operators
${c_{\alpha j}^{(\gamma)}}^\dagger(t)$. To achieve this result we
need to compute first the commutator $[c_{\alpha
j}^{(a)}(t),U_t]$, and this can be done easily by means of the
time consecutive principle, \cite{book}, and of the commutation
rules (\ref{225}): \be [c_{\alpha j}^{(a)}(t),U_t]=-i\int_0^t
[c_{\alpha j}^{(a)}(t),H_N^{(sl)}(t')]U_{t'}\,dt'= -i\int_0^t
({\rho_\alpha^j}^\dagger \Gamma_\alpha^{(a)}\delta
(t-t'))U_{t'}\,dt'=-i {\rho_\alpha^j}^\dagger \Gamma_\alpha^{(a)}
U_t.\label{226}\en Similarly we get \be [c_{\alpha
j}^{(b)}(t),U_t]=-i {\rho_\alpha^j} \Gamma_\alpha^{(b)} U_t,
\label{227}\en and, taking the adjoint of these two equations,
$$
[U_t^\dagger, {c_{\alpha j}^{(a)}}^\dagger(t)]=iU_t^\dagger
\rho_\alpha^j \overline{\Gamma_\alpha^{(a)}} \mbox{ and
}[U_t^\dagger, {c_{\alpha j}^{(b)}}^\dagger(t)]=iU_t^\dagger
{\rho_\alpha^j}^\dagger \overline{\Gamma_\alpha^{(b)}}.
$$
Going back to $\partial_tj_t(X\otimes\Id_r)$ we find that
$$
\partial_tj_t(X\otimes\Id_r)=i{\sum_{j=1}^N}\sum_{\alpha=0\pm}\mbox{\LARGE\{}\left(iU_t^\dagger
{\rho_\alpha^j}^\dagger \overline{\Gamma_\alpha^{(b)}}+{c_{\alpha
j}^{(b)}}^\dagger(t)U_t^\dagger\right)[\rho_\alpha^j,X]U_t + $$
\vspace{-3mm} $$\left(iU_t^\dagger {\rho_\alpha^j}
\overline{\Gamma_\alpha^{(a)}}+{c_{\alpha
j}^{(a)}}^\dagger(t)U_t^\dagger\right)[{\rho_\alpha^j}^\dagger,X]U_t+
U_t^\dagger [\rho_\alpha^j,X]\left(-i{\rho_\alpha^j}^\dagger
\Gamma_\alpha^{(a)} U_t+U_t c_{\alpha j}^{(a)}(t) \right)+
$$
\vspace{-3mm} $$+ U_t^\dagger
[{\rho_\alpha^j}^\dagger,X]\left(-i{\rho_\alpha^j}
\Gamma_\alpha^{(b)} U_t+U_t c_{\alpha j}^{(b)}(t)
\right)\mbox{\LARGE\}}
$$
which has to be computed on the state $\Omega_{tot}$. Therefore,
since the generator $L$ satisfies the equality
$\Omega_{tot}(\partial_tj_t(X\otimes\Id_r))=\Omega_{tot}(j_t(L(X)))$,
we  get \be
L(X)={\sum_{j=1}^N}\sum_{\alpha=0\pm}\left\{[\rho_\alpha^j,X]
{\rho_\alpha^j}^\dagger
\Gamma_\alpha^{(a)}+[{\rho_\alpha^j}^\dagger,X] {\rho_\alpha^j}
\Gamma_\alpha^{(b)}- \rho_\alpha^j [{\rho_\alpha^j}^\dagger,X]
\overline{\Gamma_\alpha^{(a)}}- {\rho_\alpha^j}^\dagger
[{\rho_\alpha^j},X]
\overline{\Gamma_\alpha^{(b)}}\right\}\label{228} \en This
expression can be made simpler if the observable $X$ is
self-adjoint ($X=X^\dagger$). In this case we have \be
L(X)=L_1(X)+L_2(X),\label{229}\en where \be
L_1(X)=\sum_{j=1}^N\sum_{\alpha=0\pm}\left\{[\rho_\alpha^j,X]
{\rho_\alpha^j}^\dagger \Gamma_\alpha^{(a)}+ h.c. \right\},
\hspace{4mm}
L_2(X)=\sum_{j=1}^N\sum_{\alpha=0\pm}\left\{[{\rho_\alpha^j}^\dagger,X]
{\rho_\alpha^j} \Gamma_\alpha^{(b)}+ h.c. \right\}.\label{230}\en

This formula will be the starting point for the analysis in the
next section.

{\bf Remark:--} Before going on, it may be interesting to stress
that, when compared with the standard perturbative approach for
the master equation for open quantum systems, \cite{martin}, the
perturbative approach based on the SLA appears quite {\em
friendly}. For instance, the so called time consecutive principle
and the new Hilbert space with ground vector $\Psi_0$ arising
after the limit $\lambda\rightarrow0$ is taken, are typical tools
of the SLA and they are essential to make many computations almost
trivial.

\section{The phase transition}

As discussed in \cite{bm,martin}, $S_N^0$ and $R_N$ are the
relevant variables whose dynamics must be considered to analyze
the phase structure of the model. These intensive operators are
both self-adjoint, so that we can use equations (\ref{229}) and
(\ref{230}) instead of (\ref{228}). As a matter of fact, in both
\cite{bm} and \cite{martin} these equations of motion are
considered only as an intermediate step to compute the equation
for $\Delta_N=\frac{1}{2}R_N^{1/2}$, which is called {\em the gap
operator}. We will see in a while that the same conclusions as in
\cite{bm,martin} can be obtained without introducing $\Delta_N$
but working directly with $R_N$ and $S_N^0$.

As a first step we compute $L(S_N^0)=L_1(S_N^0)+L_2(S_N^0)$. We
have, using (\ref{22}), (\ref{23}) and (\ref{230})
$$L_1(S_N^0)=\frac{1}{N}\sum_{j=1}^N
L_1(\sigma_j^0)=\frac{1}{N}\sum_{j=1}^N\sum_{\alpha=0,\pm}\left\{[\rho_\alpha^j,
\sigma_j^0] {\rho_\alpha^j}^\dagger
\Gamma_\alpha^{(a)}+h.c.\right\}$$ which can be written as \be
L_1(S_N^0)= \sum_{\alpha=0,\pm}\left\{(b_{\alpha +}S_N^++b_{\alpha
-}S_N^-+b_{\alpha 0}S_N^0+b_{\alpha
1}\Id_N)\Gamma_\alpha^{(a)}+h.c.\right\},\label{31}\en where
$\Id_N=\frac{1}{N}\sum_{j=1}^N\Id_j$ and the various coefficients
$\{b_{\alpha\,\gamma}\}$ have been introduced here only to stress
the fact that $L_1(S_N^0)$ is linear in the intensive operators.
As we have already mentioned before, the limit of the right hand
side of the formula exists in the strong topology restricted to a
certain family ${\mathcal F}$ of states, since all the operators
$S^\alpha_N$ converge in this topology. Therefore also the limit
of the left hand side does exist in the same topology. After some
non trivial algebra we find \be L_1(S^0):={\mathcal F}-strong
\lim_{N\rightarrow\infty}
L_1(S^0_N)=-\frac{8g^4S^0(S^+S^-)^2}{\omega^3}\left\{\Re
\Gamma_+^{(a)}\frac{\omega-g}{(\omega+gS^0)^2}+ \Re
\Gamma_-^{(a)}\frac{\omega+g}{(\omega-gS^0)^2}\right\},
\label{32}\en where $\Re \Gamma_\pm^{(a)}$ indicates the real part
of $\Gamma_\pm^{(a)}$.

The computation of $L_2(S^0):={\mathcal F}-strong
\lim_{N\rightarrow\infty} L_2(S^0_N)$ follows essentially the same
steps and produces \be
L_2(S^0)=-\frac{8g^4S^0(S^+S^-)^2}{\omega^3}\left\{\Re
\Gamma_+^{(b)}\frac{\omega+g}{(\omega+gS^0)^2}+ \Re
\Gamma_-^{(b)}\frac{\omega-g}{(\omega-gS^0)^2}\right\},
\label{33}\en so that the final result is \be
L(S^0)=-\frac{8g^4S^0(S^+S^-)^2}{\omega^3}h(S^0,S^+S^-).\label{34}\en
Here we have introduced, for brevity, the function \be
h(S^0,S^+S^-)=\Re \Gamma_+^{(a)}\frac{\omega-g}{(\omega+gS^0)^2}+
\Re \Gamma_-^{(a)}\frac{\omega+g}{(\omega-gS^0)^2}+\Re
\Gamma_+^{(b)}\frac{\omega+g}{(\omega+gS^0)^2}+ \Re
\Gamma_-^{(b)}\frac{\omega-g}{(\omega-gS^0)^2}, \label{35}\en and
we have made explicit the fact that $h$ depends on
$S^+S^-={\mathcal F}-strong \lim_{N\rightarrow\infty}S_N^+S_N^-$
via the pulsation $\omega$, see (\ref{212}).  It is interesting to
observe that the same function $h(S^0,S^+S^-)$ appears in the
computation of $L(S^+S^-):={\mathcal F}-strong
\lim_{N\rightarrow\infty} L(S_N^+S_N^-)$. Again, since
$(S_N^+S_N^-)^\dagger=S_N^+S_N^-$, we can use formulas (\ref{229})
and (\ref{230}). Here the computations are significantly harder,
but no difficulty of principle arises. As a technical tool it is
convenient to use  the fact that, in the limit
$N\rightarrow\infty$, all the intensive operators commute with all
the local operators of the system,
$\lim_{N\rightarrow\infty}[S_N^\alpha,\sigma_j^\beta]=0$, for all
$\alpha, \beta$ and $j$. Therefore we get \be
L(S^+S^-)=-\frac{16g^4(S^+S^-)^3}{\omega^3}h(S^0,S^+S^-).\label{36}\en

\vspace{3mm}

The  phase  structure of the model is now given by the right-hand
sides of equations (\ref{34}) and (\ref{36}), see
\cite{bm,martin}, and, in particular, from the zeros of the
functions \be f_1(x,y)=-\frac{8g^4xy^2}{\omega^3}h(x,y),
\hspace{5mm}f_2(x,y)=-\frac{16g^4y^3}{\omega^3}h(x,y),\label{37}\en
where we have introduced, to simplify the notation, $x=S^0$ and
$y=S^+S^-$, so that $\omega=g\sqrt{x^2+4y}$ and
$\nu=2\tilde\epsilon+gx$. In particular, the existence of a
super-conducting phase corresponds to the existence of a non
trivial zero of $f_1$ and $f_2$, \cite{bm,martin}. Due to the
definition of $f_1$ and $f_2$ it is clear that any $(x_o,y_o)$,
with $x_o\neq 0$ and $y_o\neq 0$, is such that
$f_1(x_o,y_o)=f_2(x_o,y_o)=0$ if and only if it is a zero of the
function $h$: $h(x_o,y_o)=0$. In order to find such a solution, it
is first necessary to obtain an explicit expression for the
coefficients $\Re\Gamma_\pm^{(\gamma)}$. This is easily done using
the definitions in (\ref{223}), since we get \be
\Re\Gamma_\pm^{(a)}=\frac{1}{2}\int_{-\infty}^\infty\sum_{\vec
p\in\Lambda_N}|f_m(\vec p)|^2e^{-i\tau\nu_\pm(\vec
p)}\,d\tau=\pi\sum_{\vec p\in\Lambda_N}|f_m(\vec
p)|^2\delta(\nu_\pm(\vec p)),\label{38}\en and \be
\Re\Gamma_\pm^{(b)}=\pi\sum_{\vec p\in\Lambda_N}|f_n(\vec
p)|^2\delta(\nu_\pm(\vec p)).\label{39}\en It is now almost
straightforward to recover the results of \cite{bm,martin}.
Following Buffet and Martin's original idea, we look for solutions
corresponding to $\nu=0$. This means that, because of (\ref{212}),
the value of $x=S^0$ is fixed: $x=-2\tilde\epsilon/g$. Moreover,
with this choice, $\nu_+(\vec p)=\omega-\epsilon_{\vec p}$, which
is zero if and only if $\omega=\epsilon_{\vec p}$. Also, we have
$\nu_-(\vec p)=-\omega-\epsilon_{\vec p}$, which is never zero.
For these reasons we deduce that $\Re\Gamma_-^{(\gamma)}=0$,
$\gamma=a,b$, while the sums in (\ref{38}) and (\ref{39}) for
$\Re\Gamma_+^{(\gamma)}$ are restricted to the smaller set,
$\E_N\subset\Lambda_N$, of those values of $\vec p$ such that, if
$\vec q\in\E_N$ then $\epsilon_{\vec q}=\omega$. Therefore,
recalling the expression of $m(\vec p)$ and $n(\vec p)$ in
(\ref{215bis}), we find \be
\Re\Gamma_+^{(a)}=\pi\frac{e^{\beta\omega}}{e^{\beta\omega}-1}\sum_{\vec
p\in\E_N}|f(\vec p)|^2, \hspace{4mm}
\Re\Gamma_+^{(b)}=\pi\frac{1}{e^{\beta\omega}-1}\sum_{\vec
p\in\E_N}|f(\vec p)|^2. \label{310}\en

From definition (\ref{35}), therefore, we get the following
equation
$$ \pi\frac{e^{\beta\omega}}{e^{\beta\omega}-1}\sum_{\vec
p\in\E_N}|f(\vec p)|^2\frac{\omega-g}{(\omega+gx)^2}+
\pi\frac{1}{e^{\beta\omega}-1}\sum_{\vec p\in\E_N}|f(\vec
p)|^2\frac{\omega+g}{(\omega+gx)^2}=0,
$$
or \be e^{\beta\omega}=\frac{g+\omega}{g-\omega}.\label{311}\en
This equation is the crucial one, which replaces the equation
obtained in \cite{bm,martin}, $g\tanh\left(
\frac{\beta\omega}{2}\right)=\omega$. We conclude that:

1) first of all, introducing a new variable
$\xi=\frac{\omega}{g}$, equation (\ref{311}) has a non-trivial
solution if and only if the function $g(\xi)=e^{\beta
g\xi}-\frac{1+\xi}{1-\xi}$ has a zero $\xi\neq0$. It is clear that
such a solution does exist \underline{only} if the first
derivative of $g(\xi)$, computed in $\xi=0$ is positive, i.e. when
$\beta g-2>0$. This is because $g(0)=0$ and $\lim_{\xi\rightarrow
1^-}g(\xi)=-\infty$. We recover therefore the first result of
\cite{bm,martin}, since this inequality implies the existence of a
{\em critical temperature}, $T_c:=\frac{g}{2k}$, coinciding with
that found by Martin and Buffet, such that, when $T<T_c$, a
$\tilde\xi\neq0$ does necessarily exist such that
$g(\tilde\xi)=0$, and the system is in a superconducting phase.

2) it is also possible to find the value of $y=S^+S^-$ directly
from equation (\ref{311}). However, in order to recover the same
value of the gap operator known in literature, we prefer to play a
little bit with equation (\ref{311}) in the following way:
$$
g\tanh\left(
\frac{\beta\omega}{2}\right)=g\frac{e^{\frac{\beta\omega}{2}}-e^{-\frac{\beta\omega}{2}}}
{e^{\frac{\beta\omega}{2}}+e^{-\frac{\beta\omega}{2}}}=g
\frac{e^{\beta\omega}-1}
{e^{\beta\omega}+1}=g\frac{\frac{g+\omega}{g-\omega}-1}{\frac{g+\omega}{g-\omega}+1}=\omega.
$$
This chain of equalities  shows once again how our equation
(\ref{311}) returns the same equation obtained in \cite{bm,martin}
with completely different techniques.

Vice versa, it is also straightforward to check that equation
$g\tanh\left( \frac{\beta\omega}{2}\right)=\omega$ implies
equation (\ref{311}):
$$
e^{\beta\omega}=\frac{\tanh\left(
\frac{\beta\omega}{2}\right)+1}{\tanh\left(
\frac{\beta\omega}{2}\right)-1}=\frac{\frac{\omega}{g}+1}{1-\frac{\omega}{g}}=
\frac{g+\omega}{g-\omega},
$$
and this concludes the proof of the equivalence of our approach
with that of Buffet and Martin.

\section{Conclusions and comments}

We have shown how the SLA can be successfully used to analyze the
phase structure of low temperature superconductivity analyzing a
strong coupling BCS model, considered as an open system
interacting with a bosonic thermal bath.

The procedure, which makes use of the canonical representation of
thermal states, is rather direct and is technically much simpler
than the one used in the original paper, \cite{bm}. Among the
other simplifications, for instance, a single equation $h(x,y)=0$
must be solved instead of the system $f_1(x,y)=f_2(x,y)=0$, which
is the highly transcendental system which must be solved in
\cite{bm}.

For this reason we believe that it may be worth considering other
models, still unsolved, with the simplifying strategy provided by
the SLA, since new insights may eventually  come out. For
instance, one could first  replace the bosonic reservoir with a
reservoir made of quons, \cite{quons}, in the attempt of getting a
different free time evolution for the creation and annihilation
quon operators. Following our analysis, and in particular
definition (\ref{35}) of the function $h$, this is in fact the
easiest way to get an higher value of the critical temperature
($T_c>\frac{g}{2k}$). Another possibility to achieve the same
result is to consider a second reservoir interacting with the
first one: again, in this way the free time evolution of the
bosonic operators will be different from the one considered here,
$a_j(fe^{-i\epsilon t})$. These models will be considered in a
forthcoming paper, \cite{bagnew}.

\vspace{6mm}

\noindent{\large \bf Acknowledgments} \vspace{3mm}

This work is partly supported by MURST.

\newpage

\appendix
\renewcommand{\theequation}{\Alph{section}.\arabic{equation}}


 \section{\hspace{-14.5mm} Appendix:  Few results on the stochastic limit}

In this Appendix we will briefly summarize some of the basic facts
and properties concerning the SLA which are used all throughout
the paper. We refer to \cite{book} and references therein for more
details.

Given an open system ${\cal S}+{\cal R}$ we write its hamiltonian
$H$ as the sum of two contributions, the free part $H_0$ and the
interaction $\lambda H_I$. Here $\lambda$ is a coupling constant,
$H_0$ contains the free evolution of both the system ${\cal S}$
and the reservoir ${\cal R}$, while $H_I$ contains the interaction
between ${\cal S}$ and ${\cal R}$. Working in the interaction
picture, we define $H_I(t)=e^{iH_0t}H_Ie^{-iH_0t}$ and the so
called wave operator $U_\lambda(t)$ which is the solution of the
following differential equation \be
\partial_t U_\lambda(t)=-i\lambda H_I(t)U_\lambda(t),
\label{a1} \en with the initial condition $U_\lambda(0)=\Id$.
Using the van-Hove rescaling $t\rightarrow \frac{t}{\lambda^2}$,
see \cite{martin,book} for instance, we can rewrite the same
equation in a form which is more convenient for our perturbative
approach, that is \be
\partial_t U_\lambda(\frac{t}{\lambda^2})=-\frac{i}{\lambda} H_I(\frac{t}{\lambda^2})U_\lambda(\frac{t}{\lambda^2}),
\label{a2} \en with the same initial condition as before. Its
integral counterpart is \be
U_\lambda(\frac{t}{\lambda^2})=\Id-\frac{i}{\lambda} \int_0^t
H_I(\frac{t'}{\lambda^2})U_\lambda(\frac{t'}{\lambda^2})dt',
\label{a3} \en which is the starting point for a perturbative
expansion, which works in the following way.

Suppose, to begin with, that we are interested to the zero
temperature situation. Then let $\varphi_0$ be the ground vector
of the reservoir and $\xi$ a generic vector of the system. Now we
put $\varphi_0^{(\xi)}=\varphi_0\otimes\xi$. We want to compute
the limit, for $\lambda$ going to $0$, of the first non trivial
order of the mean value of the perturbative expansion of
$U_\lambda(t/\lambda^2)$ above in $\varphi_0^{(\xi)}$, that is the
limit of \be I_\lambda(t)=(-\frac{i}{\lambda})^2\int_0^t dt_1
\int_0^{t_1}dt_2<H_I(\frac{t_1}{\lambda^2})H_I(\frac{t_2}{\lambda^2})>_{\varphi_0^{(\xi)}},
\label{a4} \en for $\lambda\rightarrow 0$. Under some regularity
conditions on the functions which are used to smear out the
(typically) bosonic fields of the reservoir, this limit is shown
to exist for many relevant physical models, see \cite{book}, and
\cite{bagacc,baglaser} for few recent applications to quantum many
body theory. It is at this stage that all the complex quantities
like the  $\Gamma_\alpha^{(\gamma)}$'s we have introduced in the
main body of this paper appear. We define
$I(t)=\lim_{\lambda\rightarrow 0}I_\lambda(t)$. In the same sense
of the convergence of the (rescaled) wave operator
$U_\lambda(\frac{t}{\lambda^2})$ (the convergence in the sense of
correlators), it is possible to check that also the (rescaled)
reservoir operators converge and define new operators which do not
satisfy canonical commutation relations but a modified version of
these. For instance, in Section II this procedure has produced the
operators $c_{\alpha\,j}^{(\gamma)}$ starting from $c_{\vec
p,\,j}^{(\gamma)}$. Moreover, these limiting operators depend
explicitly on time and they live in a Hilbert space which is
different from the original one. In particular, they annihilate a
vacuum vector, $\eta_0$,
 which is no longer the original one, $\varphi_0$. This is what happens, for instance, if $\varphi_0$  depends on
 $\lambda$, $\varphi_0\rightarrow \varphi_0^{(\lambda)}$, and
 considering $\eta_0$ as the following limit: $\eta_0=\lim_{\lambda\rightarrow
 0}\varphi_0^{(\lambda)}$.

It is not difficult to deduce the form of a time dependent
self-adjoint operator $H_I^{(sl)}(t)$, which depends on the system
operators and on the limiting operators of the reservoir, such
that the  first non trivial order of the mean value of the
expansion of $U_t=\Id-i\int_0^tH_I^{(sl)}(t')U_{t'}dt'$ on the
state $\eta_0^{(\xi)}=\eta_0\otimes\xi$ coincides with $I(t)$. The
operator $U_t$  defined by this integral equation is called again
the {\em wave operator}.

The form of the generator follows now from an operation of normal
ordering. More in details, we start defining the flux of an
observable  $\tilde X=X\otimes \Id_{r}$, where $\Id_{r}$ is the
identity of the reservoir and $X$ is an observable of the system,
as $j_t(\tilde X)=U_t^\dagger \tilde XU_t$. Then, using the
equation of motion for $U_t$ and $U_t^\dagger$, we find that
$\partial_t j_t(\tilde X)=iU_t^\dagger [H_I^{(sl)}(t),\tilde
X]U_t$. In order to compute the mean value of this equation on the
state $\eta_0^{(\xi)}$, so to get rid of the reservoir operators,
it is convenient to compute first the commutation relations
between $U_t$ and the limiting operators of the reservoir. At this
stage the so called time consecutive principle is used in a very
heavy way to simplify the computation. This principle, which has
been checked for many classes of physical models, \cite{book},
states that, if $\beta(t)$ is any of these limiting operators of
the reservoir, then \be [\beta(t),U_{t'}]=0, \mbox{ for all }
t>t'. \label{a5} \en Using this principle and recalling that
$\eta_0$ is annihilated by the limiting annihilation operators of
the reservoir, it is now a simple exercise to compute $<\partial_t
j_t(X)>_{\eta_0^{(\xi)}}$ and, by means of the equation
$<\partial_t
j_t(X)>_{\eta_0^{(\xi)}}=<j_t(L(X))>_{\eta_0^{(\xi)}}$, to
identify the form of the generator of the physical system.

\vspace{4mm}

Let us now consider the case in which $T>0$. In this case the
state of the reservoir is no longer given by $\varphi_0$. It is
now convenient to use the so-called {\em canonical representation
of thermal states}, \cite{book}. Using the same notation of
Section 2, any annihilator operator $a_{\vec p,j}$ can be written
as the following linear combination \be a_{\vec p,j}=\sqrt{m(\vec
p)}\,c_{\vec p,j}^{(a)}+\sqrt{n(\vec p)}\,c_{\vec
p,j}^{(b),\dagger},\label{a6}\en where $m(\vec p)$ and $n(\vec p)$
are the following two-points functions, \be m(\vec
p)=\omega_\beta(a_{\vec p,j}a_{\vec
p,j}^\dagger)=\frac{1}{1-e^{-\beta\epsilon_{\vec p}}},
\hspace{1cm}n(\vec p)=\omega_\beta(a_{\vec p,j}^\dagger a_{\vec
p,j})=\frac{e^{-\beta\epsilon_{\vec p}}}{1-e^{-\beta\epsilon_{\vec
p}}},\label{a6bis}\en for our bosonic reservoir, if $\omega_\beta$
is a KMS state corresponding to an inverse temperature $\beta$.
The operators $c_{\vec p,j}^{(\alpha)}$ are assumed to satisfy the
following commutation rules \be [c_{\vec p,j}^{(\alpha)},{c_{\vec
q,k}^{(\gamma)}}^\dagger]=\delta_{jk}\delta_{\vec p\,\vec
q}\delta_{\alpha\gamma}, \label{a7}\en while all the other
commutators are trivial. Let moreover $\Phi_0$ be the vacuum of
the operators $c_{\vec p,j}^{(\alpha)}$:
$$
c_{\vec p,j}^{(\alpha)}\Phi_0=0, \hspace{1cm}\forall \vec
p,j,\alpha.
$$
Then it is immediate to check that the results in (\ref{a6bis})
for the KMS state can be found, using these new variables,
representing $\omega_\beta$ as the following  vector state
$\omega_\beta(\cdot)=<\Phi_0,\cdot\Phi_0>$. With this GNS-like
representation it is trivial to check that both the CCR and the
two-point functions are easily recovered. This representation is
also called in \cite{book} the Fock-anti Fock representation
because of the different sign in the free time evolution of the
annihilation operators $c_{\vec p,j}^{(a)}$ and $c_{\vec
p,j}^{(b)}$. Once this representation is introduced, all the same
steps as for the situation with $T=0$  can  be repeated, and the
expression for the generator can be deduced using exactly the same
strategy.

\end{document}